\documentclass[twocolumn]{aastex631}

\usepackage{amsmath}
\DeclareUnicodeCharacter{02BC}{'} 
\usepackage[mathlines]{lineno}
\usepackage{xcolor}
\usepackage{ulem}
\makeatletter
\renewenvironment{acknowledgments}{
  \section*{Acknowledgments}
  \nolinenumbers 
}
\makeatother

\begin{document}
\title{Observational Study of Recurrent Jets: Evolution of Magnetic Flux, Current and Helicity}

\author[0009-0009-1911-399X]{Chang Zhou}
\affiliation{School of Astronomy and Space Science and Key Laboratory of Modern Astronomy and Astrophysics, Nanjing University, Nanjing 210023, People’s Republic of China; \href{mailto:guoyang@nju.edu.cn}{guoyang@nju.edu.cn}}

\author[0000-0002-9293-8439]{Yang Guo}
\affiliation{School of Astronomy and Space Science and Key Laboratory of Modern Astronomy and Astrophysics, Nanjing University, Nanjing 210023, People’s Republic of China; \href{mailto:guoyang@nju.edu.cn}{guoyang@nju.edu.cn}}

\author[0000-0001-9610-0433]{Guoyin Chen}
\affiliation{School of Astronomy and Space Science and Key Laboratory of Modern Astronomy and Astrophysics, Nanjing University, Nanjing 210023, People’s Republic of China; \href{mailto:guoyang@nju.edu.cn}{guoyang@nju.edu.cn}}

\author[0000-0002-1190-0173]{Ye Qiu}
\affiliation{Institute of Science and Technology for Deep Space Exploration, Suzhou Campus, Nanjing University, Suzhou 215163, Peopleʼs Republic of China}

\author[0000-0002-4978-4972]{M.D. Ding}
\affiliation{School of Astronomy and Space Science and Key Laboratory of Modern Astronomy and Astrophysics, Nanjing University, Nanjing 210023, People’s Republic of China; \href{mailto:guoyang@nju.edu.cn}{guoyang@nju.edu.cn}}

\begin{abstract}
We observed three recurrent blowout jets in an active regio with Atmospheric Imaging Assembly (AIA) aboard the Solar Dynamics Observatory (SDO). Using Helioseismic Magnetic Imager (HMI) data. We found that the magnetic flux of an emerging negative pole increases steadily before declining just as the jets erupt. Certain physical quantities, like the total unsigned vertical current, align with the periodicity of the jets. The differential affine velocity of the vector magnetograms reveals strong shear around the negative pole. The Doppler velocity map, calculated from the H$\alpha$ spectra observed by the Chinese H$\alpha$ Solar Explorer (CHASE), shows  upflows with large initial velocity before it can be observed by AIA. The magnetic field derived from the nonlinear force-free field (NLFFF) model suggests a topology akin to fan-spine structure, consistent with AIA images. We calculated the evolution of volumetric helicity ratio using the NLFFF model and found its phase aligns with the jet flux in AIA 171 \AA. These results suggest that recurrent jets may be triggered by the accumulation and release of energy and helicity, driven by  emergence, shearing and cancellation of photospheric magnetic field.

\end{abstract}

\keywords{Solar magnetic field(1503) --- Solar magnetic flux emergence(2000) --- Solar ultraviolet emission(1533) --- Solar magnetic reconnection(1504)}

\section{Introduction}\label{sec:1}
Solar jet is one of the transient activities in the solar atmosphere, which manifests morphologically as a collimated plasma flow from the photosphere to the solar corona. The typical size of the jets is $5\times10^{3}$--$4\times10^{5}$ km and the apparent velocity is 30--300 km~s$^{-1}$ and the mean lifetime is 10 minutes \citep{1996PASJ...48..123S,2007PASJ...59S.771S}. They can be observed in various spectral bands such as X-ray (\citeauthor{1992PASJ...44L.173S}~\citeyear{1992PASJ...44L.173S}), H$\alpha$ \citep{1973SoPh...28...95R,2018ApJ_Tianhui}, UV \citep{Panesar_2022ApJ,Koletti_2024A&A}, and extreme ultraviolet (EUV, \citeauthor{1998ApJ...508..899W}~\citeyear{1998ApJ...508..899W}; \citeauthor{1999SoPh..190..167A}~\citeyear{1999SoPh..190..167A}). Analysis of multiband observations shows that the jets can be of different sizes and include hot and cold plasma \citep{2000ApJ_Shimojo,2017A&A_Mulay,2019ApJ_Tiwari}.
\cite{2010_moore} divided X-ray jets into standard and blowout jets according to the presence of base arches and whether the base arches have sufficient shear and twist to support the open eruption. 

There is often emergence or cancelation of magnetic flux or both in the jet region \citep{2007A&A...469..331J,2012A&A...539A...7L,2016ApJ_panesar,2019ApJ_mcglasson,2022FrASS_schmieder}. According to the correlation with the emerging flux, \cite{1992PASJ...44L.173S} proposed the widely accepted emerging-flux model to explain the mechanism of X-ray jets. The model describes magnetic bipole emerging in an oblique open field region and reconnecting with the open magnetic field. \cite{2017A&A_Mulay,2020A&A...639A..22J} found some multi-temperature cases from
 Atmospheric Imaging Assembl (AIA) aboard the Solar Dynamics Observatory (SDO) and Interface Region Imaging Spectrograph (IRIS) to support this model. In numerical simulation, it was tested by \cite{1995Natur.375...42Y} using a two-dimensional (2D) magnetohydrodynamics (MHD) simulation. Furthermore, three-dimensional (3D) simulations with more complex magnetic field structure such as flux rope have been developed by \cite{2013ApJ...769L..21A} and  \cite{2013ApJ...771...20M}. However, based on the high resolution EUV images of AIA, \cite{2015_sterling} found dark feature named minifilament in the X-ray jets in solar coronal holes. They suggested the minifilament eruption triggered by emerging flux is the driver of the jets. Furthermore, cancelation also plays an important role in jets with minifilament eruption, which were later shown by \cite{2016ApJ_panesar,2017ApJ_panesar,2018ApJ_panesar}.
 \cite{2015A&A...573A.130P,2016A&A...596A..36P} proposed a new embedded-bipole model for straight and helical solar jets: an axisymmetrical fan-spine configuration with 3D magnetic null-point. The rotation of footpoint injects magnetic free energy and helicity into the magnetic structure and then the jet is triggered by kink instability. 

Sometimes, solar jets occur recurrently in the same active region. \cite{2011_yanglh_RAA....11.1229Y} compared the observational features of recurrent jets in EUV and soft X-ray and found that they have similar sizes, directions and velocities. \cite{2014A&A_ZhangQM} studied the thermodynamic properties of recurrent EUV jets and found similar bright and compact features called blobs when the jets rise, which may indicate the tearing-mode instability in small-scale solar activities. The reasons for the recurrent eruptions of jets remain diverse and subject to varying interpretations. From data of Transition Region and Coronal Explorer (TRACE), recurrent EUV jets were observed in instances where newly emerging flux of opposite polarity canceled with pre-existing magnetic flux \citep{1999_chae_ApJ...513L..75C}. With higher resolution magnetogram from Helioseismic Magnetic Imager (HMI), \cite{2015ApJ_chenjie...815...71C} and \cite{2019ApJ_Miao} found that recurrent jets are related to the magnetic flux cancellation between a moving satellite sunspot and the ambient opposite field. In some cases, half of the homologous jets in active region are accompanied by flux emergence, while the other half are due to cancellation \citep{2020ApJ_Paraschiv,2023ApJ_Yangliheng}. \cite{2013A&A...555A..19G} found the total absolute current and the AIA 171 \AA~ flux exhibit a consistent phase relationship, implying that recurrent reconnection may initiate the release of accumulated current, leading to recurrent jet eruptions. \citet{2024A&A_cai} found that magnetic reconnection due to 5-minute p-mode wave can lead to periodic jets in the chromosphere and transition region. \cite{2010A&A_Archonitis} analyzed recurrent jets by the 3D MHD simulation, in which an emerging toroidal flux tube reconnects with a pre-existing field. The recurrent jets carry away the free energy generated by emerging flux ropes and transform the unstable magnetic system into a stable state. \cite{2015ApJ_lee} simulated recurrent helical blowout jets and suggested that torsional Alfv\'{e}n waves play a crucial role in untwisting motion of the jets. 

With the improvement of instrument resolution in recent years, and the application of spectral observation and multi-angle observation, the rotating motion of jets has also gradually been studied. \cite{2009SoPh_Nistic} investigated 79 EUV coronal jets with STEREO/SECCHI (Sun Earth Connection Coronal and Heliospheric Investigation) and found that thirty-one jets exhibited a torsional motion around their
axis of propagation. Using data from SDO/AIA, \cite{2011ApJ_Shenyd} showed the unwinding of a coronal jet as it erupted, which was believed to be driven by the release of the magnetic twist stored in the preexisting arch through magnetic reconnection. In addition, Doppler velocity maps calculated using spectral data intuitively revealed the rotational velocity of jets \citep{2015ApJ_cheung,2018ApJ_Tiwari}. There are also observation about rotating motion in small scale jet-like event such as H$\alpha$ surge \citep{2004ApJ_jibben} and macrospicule \citep{2010A&A_kamio}. The relationship between rotating jets and magnetic field structure remains an interesting topic.

In this paper, we report three helical blowout EUV jets observed from 04:00 UT to 06:30 UT on 2022 April 15, in NOAA active region 13078. In Section \ref{sec:2}, we describe the observational data and outline the analysis methods employed. In Section \ref{sec:results}, we present the analysis results. Finally, we summarize our findings and make extended discussions in Section 4.

\section{observation and method} \label{sec:2}

\subsection{Instruments and Observation} \label{subsec:2.1}
SDO is a mission of National Aeronautics and Space Administration (NASA) designed to study the Sun and its dynamic behavior. Launched on 2010 February 11, SDO is equipped with advanced instruments that capture high-resolution images and data. AIA is one of the key instruments aboard SDO, which provides full-disk images of the corona and transition region with a spatial resolution of $1.5\arcsec$ and a temporal resolution of 12 seconds. AIA includes seven EUV filters: 94 \AA, 131 \AA, 171 \AA, 193 \AA, 211 \AA, 304 \AA, and 335 \AA, covering a temperature range from $6 \times 10^4$ K to $2 \times 10^7$ K. The HMI is another instrument aboard SDO that observes the photospheric magnetic field in 6173 Å spectral line
with a spatial resolution of $1\arcsec$, a temporal resolution of 45 seconds for line-of-sight (LOS) magnetogram and 720 seconds for vector magnetogram.

The Chinese H$\alpha$ Solar Explorer (CHASE) is the first solar space mission of China National Space Administration (CNSA). A scientific payload aboard the CHASE is H$\alpha$ Imaging Spectrograph (HIS). It provides full-Sun or region-of-interest spectral images from 6559.7--6565.9 Å and 6567.8--6570.6 Å with a spatial sampling of $0.52\arcsec$, a temporal sampling of 1 minute, 
and a spectral resolution of 0.024 Å. Note that the CHASE data used in this work is at binning mode and the corresponding resolutions except cadence need to be doubled.

On 2022 April 15, three recurrent jets occurred between 04:00 and 06:30 UT in the active region NOAA 13078. We use python package \texttt{aiapy} \citep{Barnes2020} to deconvolve the AIA data with the point spread function and upgrade the data from level 1 to level 1.5. Subsequently, we corrected for the effect of the Sun's rotation on the images to ensure the position of the event remained consistent with the first frame. CHASE/HIS observed the second jet from 05:13 UT to 05:39 UT. We used H$\alpha$ spectral data to calculate the Doppler velocity using the following equation:
\begin{equation}
v_d=\frac{\lambda-\lambda_0}{\lambda_0} \cdot c_0  ~,
\end{equation}
where $c_0$ is the speed of light in vacuum, $\lambda_0$ is the central wavelength of the average spectral line intensity in the field of view, $\lambda$ is the central wavelength calculated by centroid method.

\subsection{Method} \label{subsec:2.2}
\subsubsection{Physical parameters of photospheric magnetic field }
SHARPs (Space-Weather HMI Active Region Patches) are data products released by the SDO/HMI science team \citep{2014SoPh..289.3549B}. The vector magnetic field data patches have been  preprocessed including resolving
the $180^\circ$ ambiguity of the azimuthal component \citep{2009SoPh..260...83L} and reducing noise \citep{2012SoPh..275..285C}. To investigate the possible relationship of the recurrent jets and evolution of physical quantities in the active region, we calculated several related physical quantities using the SHARP data with the python package \texttt{calculate-sharpkeys.py} \citep{monica_g_bobra_2021_5131292}. 
The vertical current density can be derived from Ampère’s law:
\begin{equation}
J_z=\frac{1}{\mu_0}\left(\frac{\partial B_y}{\partial x}-\frac{\partial B_x}{\partial y}\right) ~,
\end{equation} 
where $\mu_0$ is the permeability of vacuum, $B_x$ and $B_y$ are the horizontal components of magnetic field, and the total absolute current is the integral of $ \left|J_z\right|$. The twist parameter $\alpha$ is defined as:
\begin{equation}
\alpha = \frac{J_z}{B_z} ~.
\end{equation}
 
To avoid the singularities at neutral line, we use the global twist \citep{2004_hagino,2009ApJ_Tiwari}:
\begin{equation}
\alpha_{g}=\frac{\sum J_z \cdot B_z}{\sum B_z^2} ~.
\end{equation}

\subsubsection{DAVE4VM}
\cite{2008ApJ...683.1134S} developed a differential affine velocity estimator for vector magnetograms (DAVE4VM) code to derive velocity fields in the photosphere and shearing motion of the magnetic footpoints. Consistent with previous works \citep{2021FrP.....9..224J}, the window size in the DAVE4VM is set to $19\times19$ pixels. We use SHARP vector magnetic field with a 12 minute cadence as input for DAVE4VM, and get photospheric velocity field. By removing the irrelevant field aligned plasma flow, \cite{2012ApJ...761..105L} corrects the $\boldsymbol{V}$ to $\boldsymbol{V}_{\perp}$  using following equation:
\begin{equation}
    \boldsymbol{V}_{\perp} = \boldsymbol{V} - \frac{\boldsymbol{V}\cdot \boldsymbol{B}}{B^{2}} \boldsymbol{B} ~.
\end{equation}

\subsubsection{DEM Analysis} \label{subsec:dem}
Differential Emission Measure (DEM) analysis is utilized for diagnosing the temperature and number density of particles in optical thin corona plasma, such as in CMEs and jets \citep{2012ApJ...761...62C,2023ApJ...943..180Z}. The flux $F_i$ in the $i$-th EUV channel can be expressed as:
\begin{equation}
F_i=\int_0^{\infty} K_i(T) \operatorname{DEM}(T) d T + \Delta F_i ~,
\end{equation}
where $K_i(T)$ is the temperature response function and $\Delta F_i$ is the synthesis of background, instrumental and statistical error. ${DEM}(T)$ is the differential emission measure of plasma along the line-of-sight written as ${DEM}(T) dT = \int_0^{\infty} {n(T)}^2 dh$, where $n(T)$ is the electron number density at temperature $T$. In this study, We employ the regularization method \citep{2012A&A_hannah} for DEM inversion, using data from 6 AIA EUV channels: 94 \AA, 131 \AA, 171 \AA , 193 \AA , 211 \AA, and 335 \AA. 
In addition, we calculate the weighted average temperature by 
\begin{equation}
\bar{T}=\frac{\int \operatorname{DEM}(T) \times T d T}{\int \operatorname{DEM}(T) d T}
\end{equation}
The integral range of temperature is $5.5 \leqslant \log T \leqslant 7.5$.

\subsubsection{Nonlinear Force-free Field (NLFFF) Modeling}
The NLFFF model is a widely used model for deducing 3D magnetic field in the solar atmosphere. 
\cite{2016_guo_b,2016_guo_a} developed a magneto-frictional method for NLFFF model in  the Message Passing Interface Adaptive Mesh Refinement Versatile Advection Code (MPI-AMRVAC, \citeauthor{KEPPENS_2012}~\citeyear{KEPPENS_2012},  \citeauthor{xia_2018}~\citeyear{xia_2018}, \citeauthor{keppens_2023}~\citeyear{keppens_2023}). We use the $Br$, $Bt$, and $Bp$ data from hmi.sharp\_cea\_720s as the bottom boundary conditions for the magneto-frictional method. 
The computational domain spans a size of $240\times 240$ cells with a spatial resolution of 0.03 degree in heliocentric angle. Then, we remove the
magnetic force and torque of the vector magnetic field of bottom boundary using the preprocessing method of \cite{2006_wiegelmamn}. Subsequently, we computed the potential field and relaxed it to force-free state based on the vector magnetic field on the bottom boundary.

Furthermore, with the sequential 3D magnetic field data during the jet eruption process, we can analyze the relationship between 
helicity evolution and the jet eruption. 

For the magnetic helicity, which is not gauge-invariant in coronal open field, we use the relative magnetic helicity \citep{1984_berger}:
\begin{equation}
H_R = \int_V (\boldsymbol{A} + \boldsymbol{A}_p)\cdot(\boldsymbol{B}-\boldsymbol{B}_p) dV ~,
\end{equation}
where $\boldsymbol{B}$ is the result of NLFFF model, $\boldsymbol{B}_p$ is corresponding potential field, $\boldsymbol{A}$ and $\boldsymbol{A}_p$ are the vector potentials satisfying $\boldsymbol{B} = \nabla \times \boldsymbol{A}$ and $\boldsymbol{B}_p = \nabla\times \boldsymbol{A}_p$. 
\cite{2012_Valori} proposed a finite volume method to calculate $\boldsymbol{A}$ and $\boldsymbol{A}_p$ within a finite rectangular volume $V=\left[x_1, x_2\right] \times\left[y_1, y_2\right] \times\left[z_1, z_2\right]$ under DeVore-GV gauge as 
\begin{equation}
\boldsymbol{A}=\boldsymbol{b}+\hat{z} \times \int_z^{z_2} \boldsymbol{B} d z^{\prime} ~,
\label{e:A}
\end{equation}
where $\boldsymbol{b}$ is an integration vector whose special solution $\boldsymbol{\Bar{b}}$ satisfy
\begin{equation}
 \bar{b}_z=0~,
\end{equation}
\begin{equation}
 \bar{b}_x=-\frac{1}{2} \int_{y_1}^y B_z\left(x, y^{\prime}, z=z_2\right) \mathrm{d} y^{\prime} ~,
\end{equation}
\begin{equation}
 \bar{b}_y=\frac{1}{2} \int_{x_1}^x B_z\left(x^{\prime}, y, z=z_2\right) \mathrm{d} x^{\prime} .
\end{equation}
Potential field $\boldsymbol{B}_p$ and vector potential $\boldsymbol{A}_p$ 
 also satisfy equation (\ref{e:A}). We realized these computational process in Python.

The gauge invariance requires the normal component of $\boldsymbol{B}$ and $\boldsymbol{B}_p$ in six boundary to satisfy $\hat{\boldsymbol{n}} \cdot \boldsymbol{B}=\hat{\boldsymbol{n}} \cdot \boldsymbol{B}_p$, where $\hat{\boldsymbol{n}}$ is unit normal vector. Using scalar potential $\phi$, $\boldsymbol{B}_p$ can be expressed as $\boldsymbol{B}_p = \nabla \phi$. Therefore, the major computation of this method is solving 3D Laplace equation $\Delta \phi = 0$ under Neumann boundary conditions $\frac{\partial \phi}{\partial \hat{n}}=\hat{\boldsymbol{n}} \cdot \boldsymbol{B}$. Refer to previous work, we used the Helmholtz solver in the proprietary Intel® Mathematical Kernel Library (MKL) to solve this Laplace problem.

\cite{1999_berger} divided $H_R$ into two gauge-invariant quantities:
\begin{equation}
H_J = \int_V (\boldsymbol{A} - \boldsymbol{A}_p)\cdot(\boldsymbol{B}-\boldsymbol{B}_p) dV ~,
\end{equation}
\begin{equation}
H_{PJ} = 2\int_V \boldsymbol{A}_p\cdot(\boldsymbol{B}-\boldsymbol{B}_p) dV ~,
\end{equation}
where $H_J$ is the current-carrying helicity and $H_{PJ}$ is the volume-threading helicity between the potential field and
the current-carrying field. The gauge-invariant $H_R$, $H_J$, and $H_{PJ}$ allow us to compare them at different time in the same volume. We also calculate the so-called helicity radio $H_J/H_R$, which is an important indicator in eruptive activities of solar active region \citep{2017A&A_pariat}.

\section{results} \label{sec:results}
\begin{figure*}
    \centering
    \includegraphics[width=\textwidth,clip]{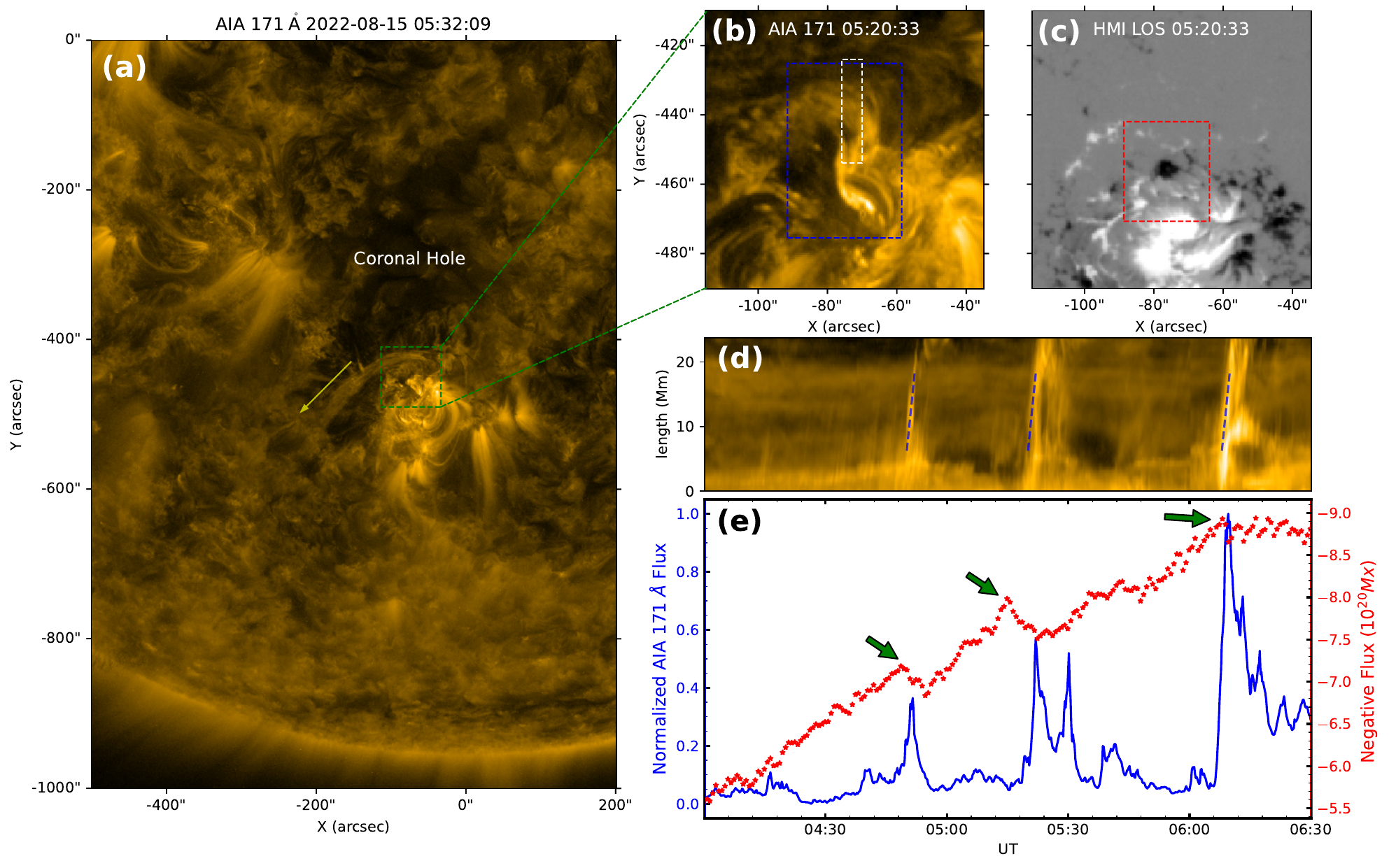}
    \caption{(a) AIA 171 \AA\ image of the jet region and nearby region. Yellow arrow marks the trajectory of the jet out of the active region. (b) AIA 171 \AA\ image of the jet source region. Blue box marks the region for calculating the EUV flux and white box marks the region for plotting the time-distance map. (c) HMI LOS magnetogram of the jet source region. Red box marks the region for calculating the negative flux. The movie of panel (b) and panel (c) can be found online. (d) Time-distance map of the white dashed line box in panel (b). Since the jet has a certain width, we average the white box in the x direction to draw the map. (e) Evolution of the normalized AIA 171 \AA~ 
    flux (blue) and negative magnetic flux (red) from 04:00--06:30 UT. }
    \label{fig:flux}
\end{figure*}

Figure \ref{fig:flux} shows the basic information about the recurrent jets. The active region is located in the mid-latitudes of the southern hemisphere, with a coronal hole situated in the north. Figure \ref{fig:flux}(c) shows the LOS magnetogram of the active region. The negative pole within the red box corresponds directly to the footpoints of the three jets. Figure \ref{fig:flux}(d) displays the time-distance map of the jets in their initial stage, whose value is the average of transverse 5 pixels. The intensity of the jets grows one by one and their mean velocity is 115 km~s$^{-1}$. Figure \ref{fig:flux}(e) presents the evolution profiles of the normalized AIA 171 \AA~ flux and negative magnetic flux. Notably, the absolute value of the negative flux increases steadily and then declines just as the jets erupt. We suppose that the cancellation is the trigger of the eruptions and the accumulation of the emerging flux enhances the intensity of eruptions.

\subsection{Erupting process of the jets}
\begin{figure*}
    \centering
    \includegraphics[width=0.9\textwidth,clip]{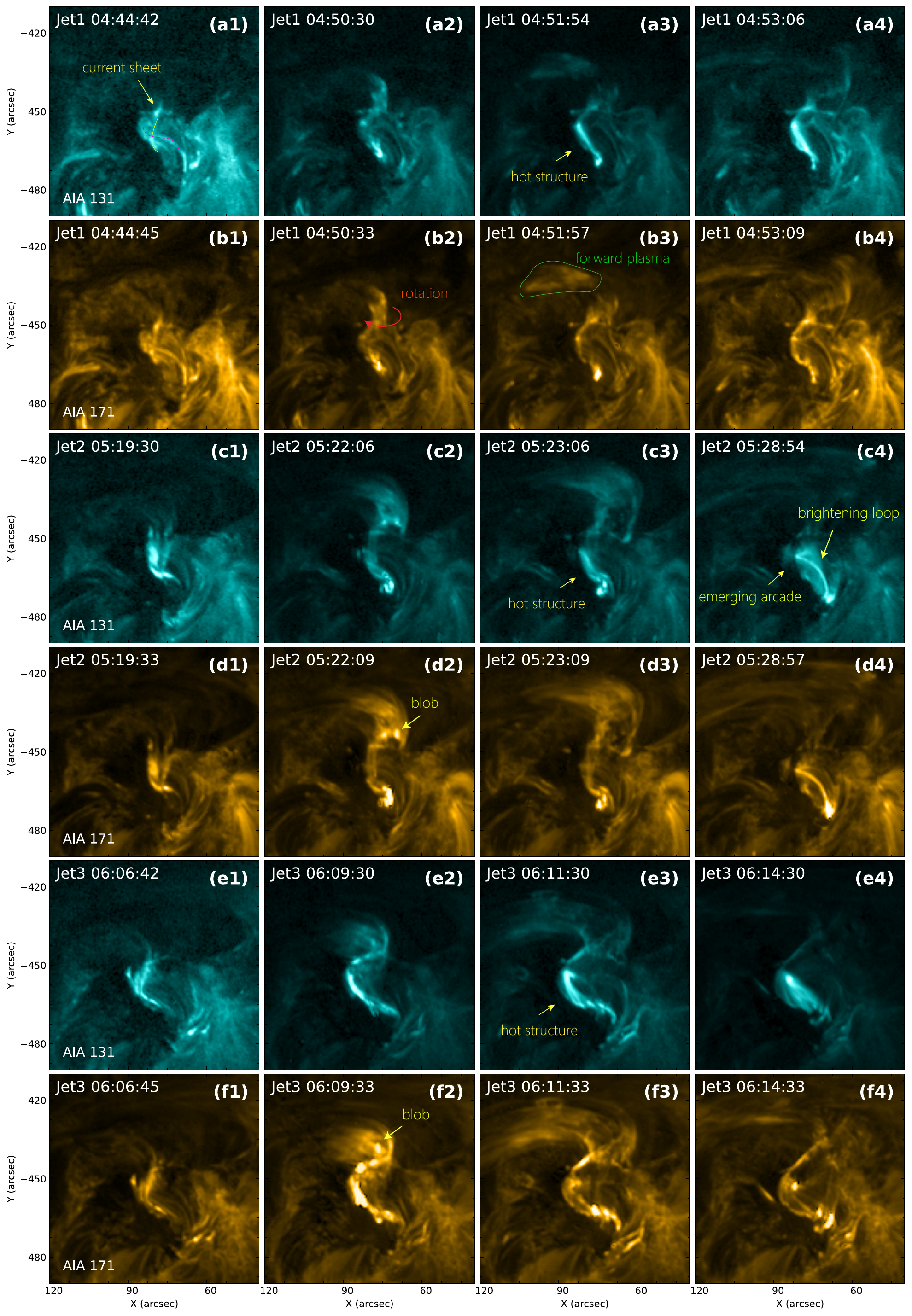}
    \caption{AIA 131 and 171 \AA\
    images of the three jets. Some special structures have been marked.}
    \label{fig:jet_all}
\end{figure*}

Figure \ref{fig:jet_all} shows the evolution of the three jets in AIA 131 and 171 \AA. The three homologous jets exhibit similar shapes and rotations. Unlike typical jets with collimated, beam-like ejecting plasma flows \citep{2021RSPSA.47700217S}, these jets have a certain width and undergo an almost 120$^\circ$ change in the ejection direction. The yellow arrow in Figure \ref{fig:jet_all}(a1) indicates the current sheet before ejection, where magnetic reconnection brightens the loop marked by yellow arc. Figure \ref{fig:jet_all}(b2) describes the left-handed rotation helix. Figures \ref{fig:jet_all}(a3), \ref{fig:jet_all}(c2), and \ref{fig:jet_all}(c3) show the hot structure in the 131 Å wavelength band, whose direction becomes more horizontal over time. In Figure \ref{fig:jet_all}(b3), we find that it is not consecutive from the forward plasma to the subsequent jet. This also occurs during the jet 2 and 3 processes, though less prominently. This indicates that these are blowout jets propelled by ejection of helical magnetic structure rather than standard jets caused by simple reconnection. Figure \ref{fig:jet_all}(c4) shows an emerging dark arcade and a brightening loop. Then, the plasma in the arcade is heated and becomes invisible. Additionally, Figure \ref{fig:jet_all}(d2) and \ref{fig:jet_all}(f2) show blobs in jet 2 and jet 3, which are also observed by \cite{2014A&A_ZhangQM} in another case and indicate tearing-mode instability in current sheets.

\begin{figure*}
    \centering
    \includegraphics[width=\textwidth,clip]{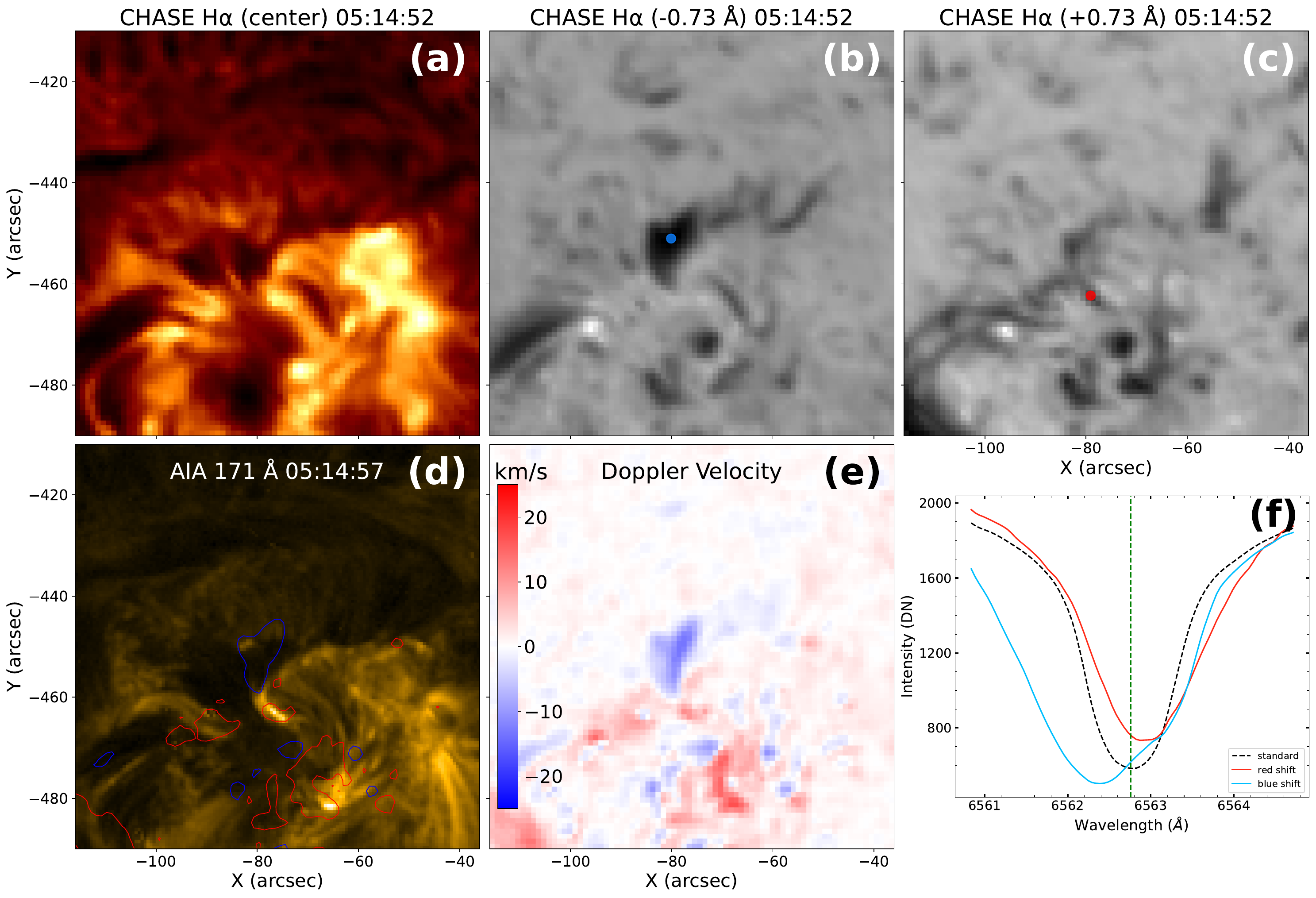}
    \caption{(a--c) CHASE H$\alpha$ images in the central wavelength, -0.73 \AA, and +0.73 \AA\ of the active region before the second jet. (d) AIA 171 Å image. The red and blue contours show Doppler velocity with levels of +5 and -5 km~s$^{-1}$. (e) Doppler velocity map. (f) H$\alpha$ profiles in different regions. The black dashed line is the average H$\alpha$ line profile, the blue solid line is the profile of blue point in panel (b), and the red solid line is the profile of red point in panel (c).}
    \label{fig:chase}
\end{figure*}

Figure \ref{fig:chase} shows the CHASE H$\alpha$ maps and the contemporaneous AIA 171 \AA~ map at the onset of the second jet. In Figure \ref{fig:chase}(b), we can see strong absorption in the H$\alpha$ blue wing, and the representative spectral line at the blue point shown as the blue profile in  Figure \ref{fig:chase}(f). Similarly,  Figure \ref{fig:chase}(c) shows absorption in the H$\alpha$ red wing, and the representative spectral line at the red point is the red profile in Figure \ref{fig:chase}(f). The Dopplor velocity contours in Figure \ref{fig:chase}(d) indicates that there were outflow and inflow of plasma before they become visible in EUV band. This suggests that pre-existing magnetic reconnection had already accelerated the plasma before it was heated to temperatures detectable in the EUV band.

\begin{figure}
    \centering
    \includegraphics[width=0.5\textwidth,clip]{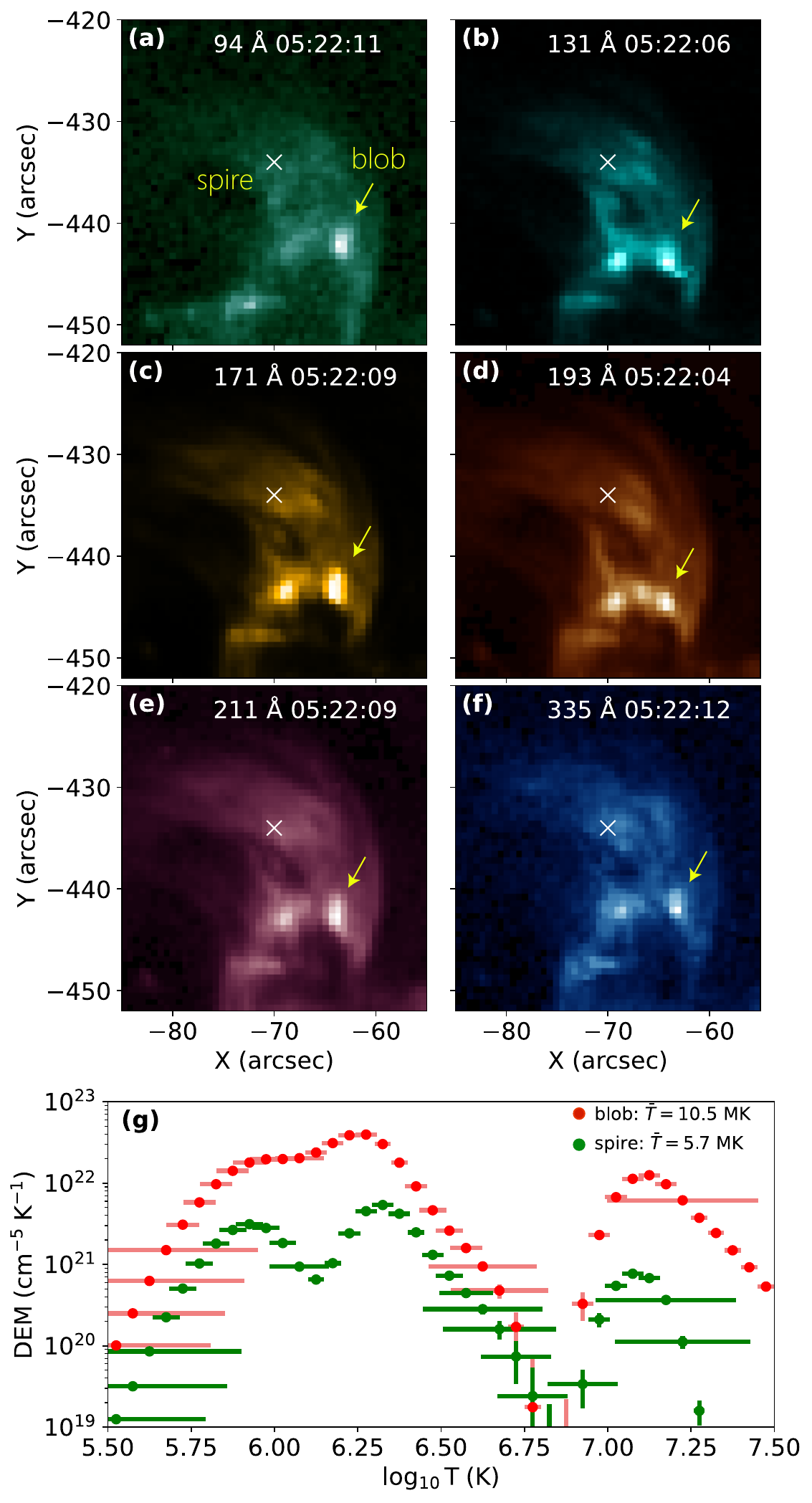}
    \caption{Images of the second jet in 6 AIA EUV channels and the DEM profile of spire and blob.}
    \label{fig:dem2}
\end{figure}

\begin{figure}
    \centering
    \includegraphics[width=0.5\textwidth,clip]{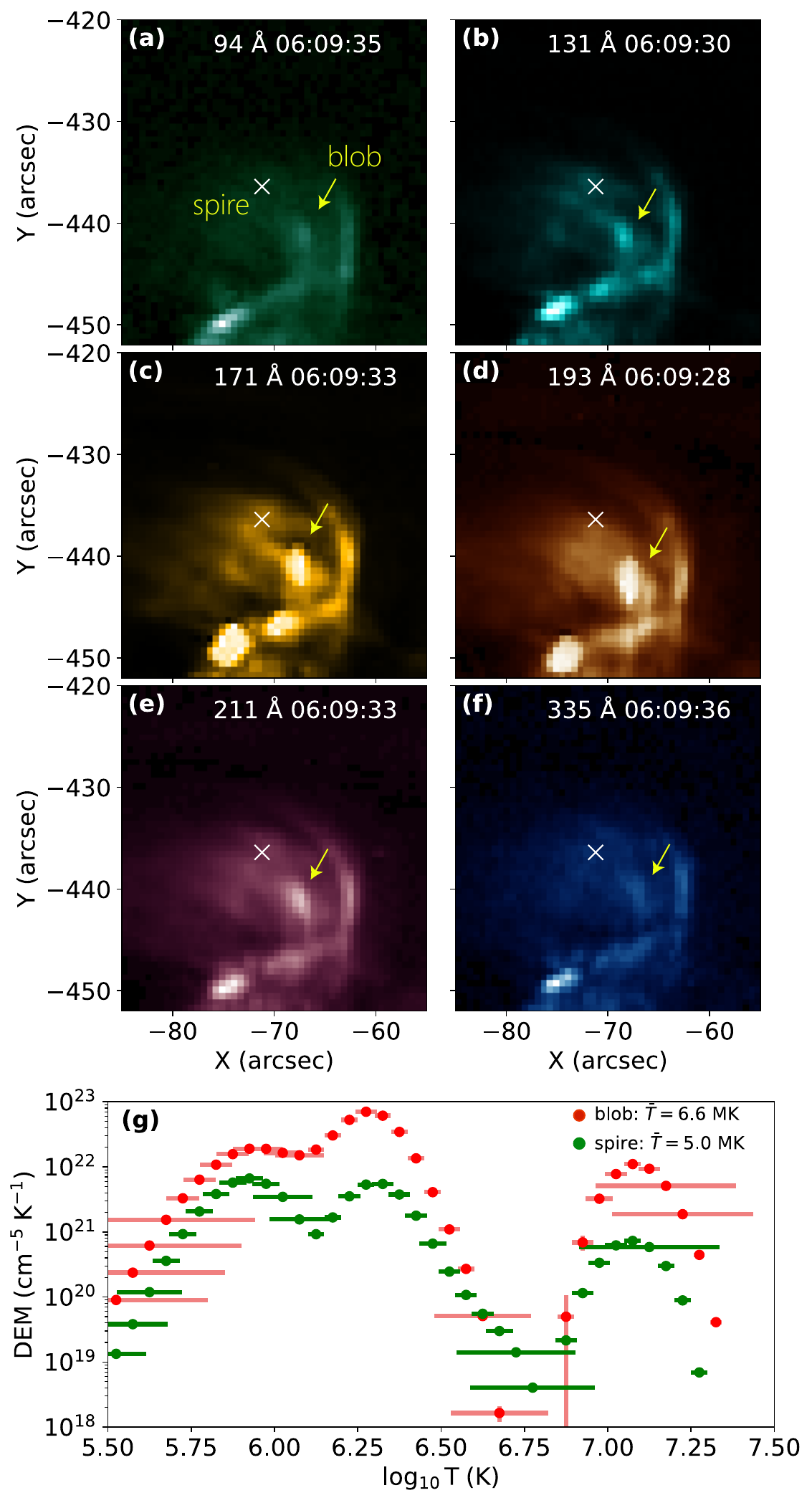}
    \caption{Images of the third jet in 6 AIA EUV channels and the DEM profile of spire and blob.}
    \label{fig:dem3}
\end{figure}

We perform a DEM analysis to the spire and blob of jet 2 and jet 3. In these two jets, the blobs exhibit a lifetime of approximately 30--60 s.  Figures \ref{fig:dem2} and \ref{fig:dem3} show the snapshots of the jets in 6 EUV channels. The blobs appear as oval bright points and there are two blobs in jet 2 at the same time. The DEM analysis indicates that thermodynamic property is similar for the jets and the blobs have higher temperature and density compared with the spire. 

\subsection{Vector magnetic field of the jets}
\begin{figure*}
    \centering
    \includegraphics[width=\textwidth,clip]{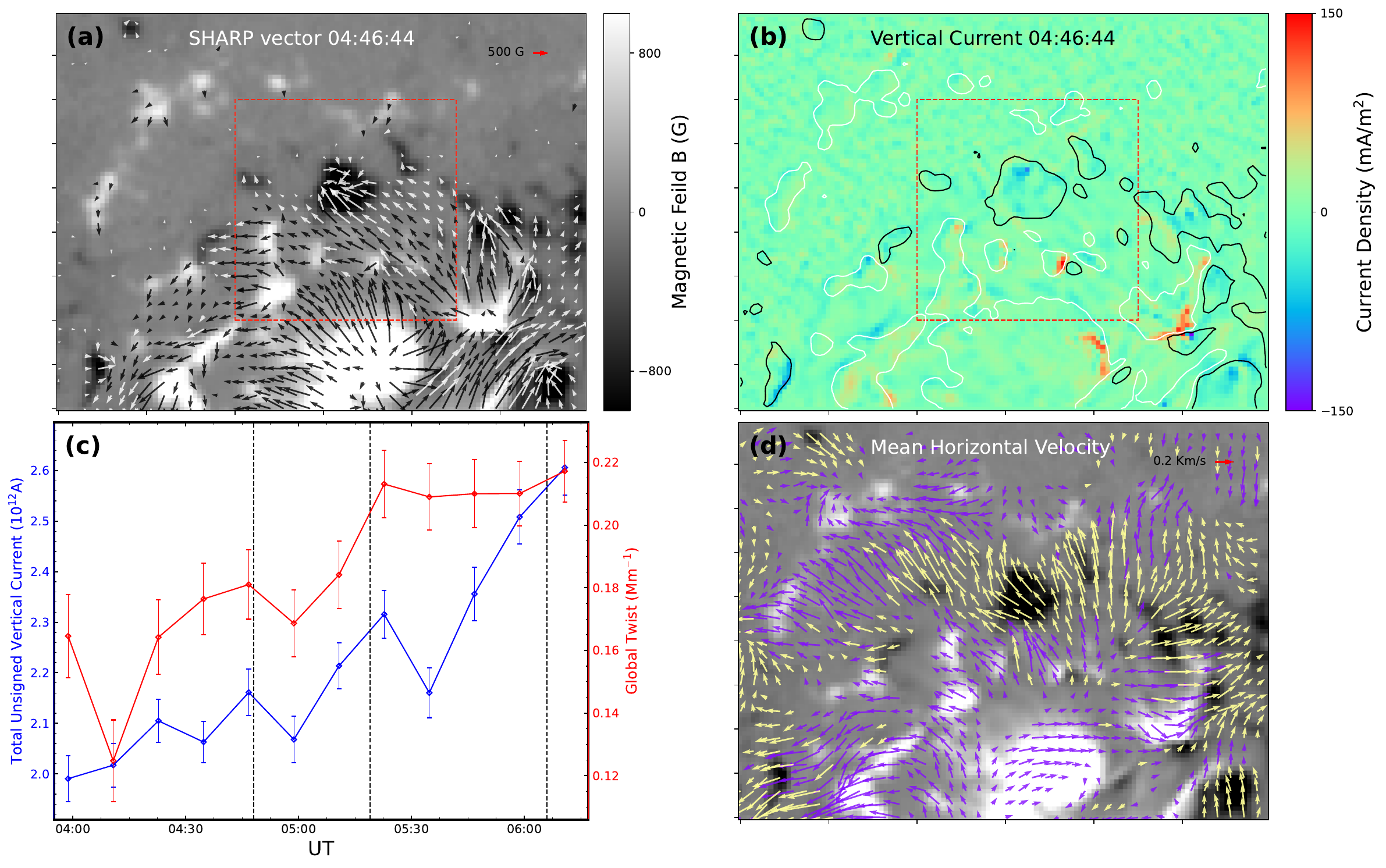}
    \caption{(a) SHARP vector magnetic field of the second jet. The background is LOS magnetogram and the black/white arrows show horizontal components located on positive/negative polarities of the background. Red box marks the active region for calculating the integral of unsigned current and  \text{global} twist parameter. (b)  Vertical current density of the active region. White/black profile are the contour of LOS magnetogram whose levels are +200/-200 Gauss. (c) Total unsigned vertical current (blue) and  \text{global} twist parameter (red) of the active region in red box of panel (a). Black vertical dashed lined mark the time when the jets erupt. (d) Mean horizontal velocity field derived from DAVE4VM. The purple/yellow arrows indicate the velocity on positive/negative polarities of the background. The background is the same LOS magnetogram as (a).}
    \label{fig:vector_m}
\end{figure*}

Figure \ref{fig:vector_m}(a) shows the vector magnetic field of the active region. In the emerging negative polarity and around the polarity inversion line, the magnetic field exhibits strong shear. Figure \ref{fig:vector_m}(b) shows the strong vertical current in the corresponding location. In Figure \ref{fig:vector_m}(c), the phase of the total absolute vertical current aligns with the phase of the jet explosion, corroborating the findings of \cite{2013A&A...555A..19G}. However, because the SDO is obscured by the Earth, the data is missing after 06:12 UT. Therefore, we cannot capture the complete profile of the third jet. As a quantity that is normalized with magnetic field intensity, the  global twist parameter indicates how twisted the magnetic field is in the photosphere. Similar to the vertical current, its phase also aligns with eruptions. 

We calculated the velocity field from 04:00--06:00 UT by DAVE4VM method. Figure \ref{fig:vector_m}(d) shows mean horizontal velocity field in the photosphere overlaid on the magnetogram. During the entire time period, the negative magnetic field that we focus on moves to northeast, contrary to the positive polarity below. This continuous shearing motion in the photosphere may facilitate the formation of coronal shear magnetic structure in the corona. 

\subsection{Magnetic topology and helicity}
\begin{figure*}
    \centering
    \includegraphics[width=\textwidth,clip]{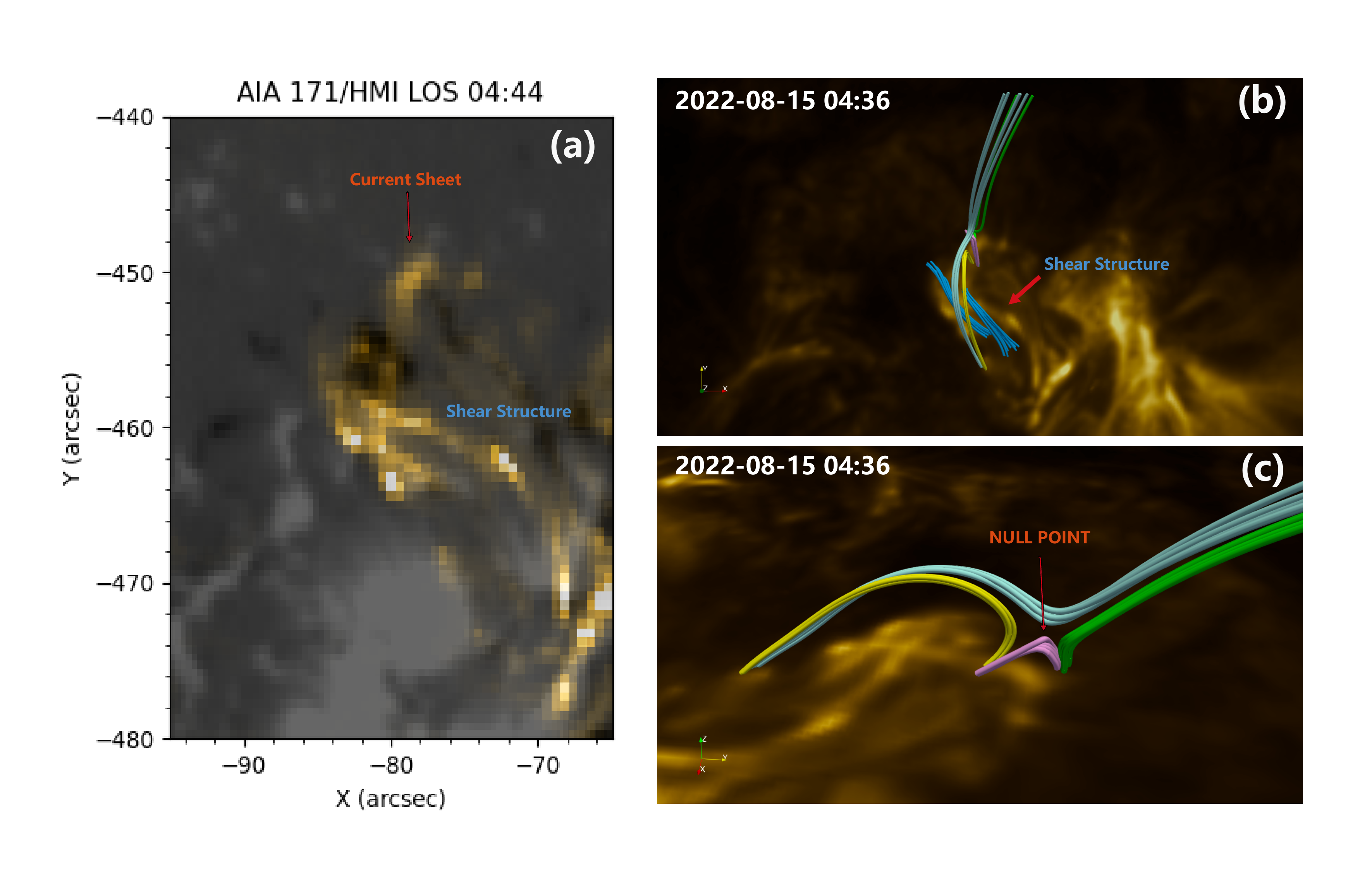}
    \caption{(a) An overlay of AIA 171 \AA\ image onto HMI LOS magnetogram with a certain level of transparency before the first eruption. (b, c) Extrapolated
magnetic field lines over the AIA 171 \AA\ image from top view and side view. The blue lines are the shear structure under the magnetic arcade. The red arrow points to null point existing in the X-type reconnection site.}
    \label{fig:nlfff}
\end{figure*}

Based on the NLFFF model, we calculate the 3D magnetic field from 04:24 UT to 06:12 UT. Figure \ref{fig:nlfff} depicts the contrast of observation and simulated magnetic topology before the complete eruption of the first jet. In Figure \ref{fig:nlfff}(a), the red arrow illustrates a  brightening structure resembling a current sheet. A brightening arch connects the current sheet to another footpoint. An S-type structure is visible under the arch. In Figures \ref{fig:nlfff}(b) and \ref{fig:nlfff}(c), there is a typical X-type reconnection topology. The light blue and green lines correspond to the initial direction of the jets. The projection of the null point and the blue magnetic lines align with the current sheet and the S-type structure in Figure \ref{fig:nlfff}(a) respectively. Therefore, we propose that the jet erupting process follows the blowout jet model \citep{2013ApJ...769L..21A}: the emergence of magnetic flux triggers reconnection at the null point, which opens the restrained arch above the shear structure. Subsequently, the shear structure reconnects with nearby field lines, ultimately ejecting as a jet and leading to a reconstruction of the magnetic topology, which also leads to the cancellation of the flux.

\begin{figure*}
    \centering
    \includegraphics[width=\textwidth,clip]{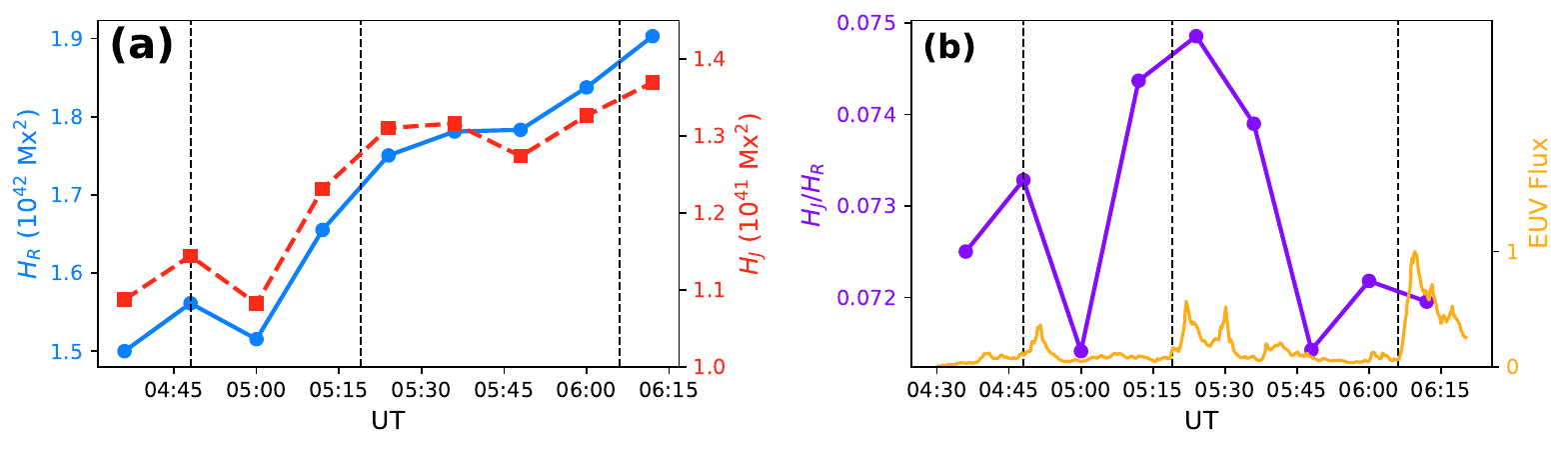}
    \caption{(a) The relative helicity (blue solid line) and the current-carrying helicity (red dashed line). (b) The helicity ratio $H_J$/$H_R$ (purple solid line with dotts) and AIA 171 \AA\ flux (orange solid line) as a contrast. The three vertical dashed lines in (a) and (b) mark the time when jets started.}
    \label{fig:H}
\end{figure*}

\cite{2020_thalmann} demonstrated the reliability of relative helicities calculated from the NLFFF. Figure \ref{fig:H}(a) shows the evolution of the helicity. The relative helicity increases continuously, and the current-carrying helicity, which is non-potential, shows some correlation with the jets. Figure \ref{fig:H}(b) shows a high degree of correlation between helicity ratio and the EUV flux, despite a 12 minute uncertainty in timing. Many previous studies \citep{2020_thalmann,2021A&A_Gupta,2023A&A_pariat} suggest the helicity ratio $H_J$/$H_R$ is important for solar activities. In the MHD simulation for single coronal jet by \cite{2023A&A_pariat}, the $H_J$/$H_R$ reaches its highest value at the time of the jet eruption and then rapidly decreases. This matches our results and indicates the importance of helicity ratio not only in single eruption but also in recurrent eruptions.

\section{Discussion}\label{discussion}
Different with common collimated jets moving along open field, the three jets change their direction after ejection and moves along closed magnetic field lines connecting the active region to the polar region. In our NLFFF extrapolation with small field of view, we can get magnetic structure similar to AIA observation and the direction of the spine line is north. If we select a larger field of view for NLFFF extrapolation or potential-field source-surface (PFSS) extrapolation, we cannot get magnetic lines in accordance with the entire ejection trajectory. It may be due to the weak surrounding magnetic field and the limitations of the extrapolation method. 

In many 3D simulations about jet \citep{2013ApJ...769L..21A,2013ApJ...771...20M,2017Natur_wyper}, a twisted magnetic flux rope is placed below the photosphere. Then, the magnetic flux rope emerges and reconnect with ambient field. In the minifilament model \citep{2015_sterling}, the dark plasma in X-ray jet is known as profile of magnetic flux rope. \cite{2017ApJ_zhuxiaoshuai} studied a blowout jet by forced field extrapolation. They found that the magnetic flux rope existed in the source region of the jet before the eruption and disappeared after the eruption. Does the magnetic flux rope necessary for blowout jet? In our case, Figure \ref{fig:nlfff} shows that the shear structure in AIA 171 \AA~corresponds well to the structure in NLFFF extrapolation. We also compared the results before the eruption and after the eruption, and found the overall magnetic structure did not change much. The change of the magnetic topology is reflected in the change of helicity in Figure \ref{fig:H}. Can this magnetic topology erupt by tether-cutting reconnection? This need to be verified by subsequent MHD simulation studies.

In previous studies on recurrent jets, researchers often paid more attention to moving of satellite sunspot or flux cancellation and emergence \citep{2015ApJ_chenjie...815...71C,2019ApJ_Miao,2020ApJ_Paraschiv}. Moreover, \cite{2013A&A...555A..19G} studied the repetitively accumulated currents and \cite{2023ApJ_Yangliheng} investigated more parameters such as Poynting flux and helicity injection rate across the photosphere. On this basis, we not only discuss the influence of these photospheric surface parameters on the recurrent jets, but also study the changes of the specific magnetic topology and magnetic helicity during the recurrent jets by NLFFF extrapolation. Previously,  these parameters were often used to study the evolution of active regions or flares.
\cite{2011ApJ_ravindra} analysed the relationship between the net, dominant, and non-dominant current with X-class flares. \cite{2023ApJ_liuyang} investigated the relative magnetic helicity of 21 X-class flares by NLFFF extrapolations and found significant decrease after eruptive flares but no clear change after confined flares. \cite{2021A&A_Gupta} discovered the CME-associated flares used to occur on the AR with helicity ratio $\langle | H_{J}\left|/\left|H_{V}\right|\right\rangle>0.1$. In terms of morphology and eruption process, blowout jet and CME have certain similarities, which can be described by the universal model for solar eruptions \citep{2017Natur_wyper}. Therefore, it is reasonable for us to study these blowout jets by magnetic helicity. Differently, according to the intensity of eruption, the change of magnetic helicity in blowout jets is not very significant compared with CMEs. So the quantity $\left | H_{J}\right|/\left|H_{V}\right| $ that expresses the relative non-potential property of magnetic field is more effective. \cite{2023A&A_pariat} found  $\left | H_{J}\right|/\left|H_{V}\right| $ decreases rapidly during the jet eruption in single jet-producing simulation. We confirmed this conclusion and extended it to recurrent jets.  
The actual value is about 0.07, which is similar to the value of the case of \cite{2023ApJ_yufu} but do not reach the level to produce CME-associated flares \citep{2021A&A_Gupta}.

The occurrence of recurrent jets represents not only a process including repeated change of magnetic topology, but also a process of repeated release and accumulation of energy in the corona. In the long-standing fan-spine structure, flux emergence continuously transports energy and plasma from the photosphere to the corona, trigering reconnections as a long-term energy release process, which aligns with recent observations and simulations \citep{2023NatCo_cheng,2024NatAs_lu} where persistent reconnections in the current sheet can effectively heat the corona. In this sense, not only the eruptive reconnection but also the hot structure after the ejection should be followed with interests. Moreover, our work reveals that, in addition to flux emergence, the photospheric shearing is also important for energy and helicity injection. The shearing motion can accelerate the energy release process by changing the standard jet to blowout jet. This provides a new perspective on understanding the coronal jets and coronal heating problem.

\section{Summary}\label{sec:summary}
In this paper, We studied three recurrent jets at NOAA active region 13078. The major results are as follows:
\begin{enumerate}
    \item \hangindent=2em Nature of Jets: These jets are blowout jets. They were ejected to the north with an initial velocity about 115 km~s$^{-1}$ and experienced significant directional changes. Blobs were observed in the second and third jets, exhibiting higher temperature and density  compared to the spire. Doppler velocity measurements of the second jet shows plasma outflows before the jet spire became visible in EUV observations. 
    \item \hangindent=2em Magnetic flux: The jets occurred on an emerging negative pole, whose flux showed an overall increasing trend with a notable decline during the jet eruption. This suggests that flux emergence and cancellation are crucial for the formation and triggering of these jets.
    \item \hangindent=2em Shear of Magnetic Field and Velocity Field: The vector magnetic field and the velocity field of the active region displayed strong shear. The phase of the total unsigned vertical current was consistent with the peak of the jets. 
    \item \hangindent=2em NLFFF Analysis: NLFFF revealed an X-shaped magnetic structure with a null point corresponding to the current sheet brightening before the eruption. The shear structure below the magnetic arch matched the AIA observations. The phase of helicity ratio calculated with NLFFF aligned with eruptions of the recurrent jets.
\end{enumerate}

\begin{acknowledgments}
C.Z., Y.G., G.Y.C and Y.Q. were supported by the National Key R\&D Program of China (2022YFF0503004, 2021YFA1600504, and 2020YFC2201201) and NSFC (12333009). The numerical computation was conducted in the High Performance Computing Center (HPCC) at Nanjing University.
\end{acknowledgments}

\bibliography{sample631}{}
\bibliographystyle{aasjournal}



\end{document}